\documentclass{article} 

\usepackage{amscd,amsmath,amssymb,amsfonts,latexsym,mathrsfs,amsthm}
\usepackage{graphicx} 
\usepackage{graphics,epsfig}

\usepackage{tabularx}
\usepackage{bbm}

\usepackage{algorithm}
\usepackage{algorithmic}
\usepackage{capt-of}
\usepackage{sectsty}
\usepackage{natbib}
\usepackage[english]{babel}
\usepackage{amsmath}
\usepackage{amsfonts}
\usepackage{amssymb,amsthm}
\usepackage{bm}
\usepackage{mathrsfs}
\usepackage{subfigure}
\usepackage[usenames]{color}

\usepackage{slashbox}
\usepackage{flushend}
\usepackage{multirow}
\usepackage{makecell}

\makeatother

\begin{document}

\title{Adaptive Rejection Sampling with  fixed number of nodes}
\author{L. Martino$^\star$, F. Louzada$^\star$ \\
$^\star$ Institute of Mathematical Sciences and Computing, \\
Universidade de S\~ao Paulo,  Brazil.
}



\date{}

\newtheorem{theo}{Theorem}
\newtheorem{defi}{Definition}
\newtheorem{coro}{Corollary}
\newtheorem{prop}{Proposition}[section]
\theoremstyle{remark}
\newtheorem*{rem}{{\bf Remark}}
\newtheorem{lemma}{Lemma}
\newtheorem{proper}{Property}[section]
\newtheorem{algo}{Algorithm}

%
\maketitle

\begin{abstract}	
The adaptive rejection sampling (ARS) algorithm is a universal random generator for drawing samples efficiently from a univariate log-concave target probability density function (pdf). ARS generates independent samples from the target via rejection sampling with high acceptance rates. Indeed, ARS yields a sequence of proposal functions that converge toward the target pdf, so that the probability of accepting a sample approaches one. However, sampling from the proposal pdf becomes more computational demanding each time it is updated. In this work, we propose a novel ARS scheme, called Cheap Adaptive Rejection Sampling (CARS), where the computational effort for drawing from the proposal remains constant, decided in advance by the user. For generating a large number of desired samples, CARS is faster than ARS. 
\newline
\newline
{\bf keyword:} Monte Carlo methods; Rejection Sampling; Adaptive Rejection Sampling
\end{abstract}


\section{Introduction}
Random variate generation is required in different fields and several applications, such as Bayesian inference and simulation of complex systems \citep{Devroye86,Hormann03,Robert04,AlmostRejLess}.
 Rejection sampling (RS) \citep[Chapter 2]{Robert04} is a universal sampling method which generates independent samples from a target probability density function (pdf).  The sample is either accepted or rejected by an adequate test of the ratio of the two pdfs. However, RS needs to establish analytically a bound for the ratio of the target and proposal densities. 

Given a target density, the adaptive rejection sampling (ARS) method \citep{Gilks92,Gilks92derfree} produces jointly both a suitable proposal pdf and the upper bound for the ratio of the target density over this proposal. Moreover, the main advantage of ARS is that ensures high acceptance rates, since ARS yields a sequence of proposal functions that actually converge toward the target pdf when the procedure is iterated. The construction of the proposal pdf is obtained by a non-parametric procedure using a set of support points (nodes), with increasing cardinality. When a sample is rejected in the RS test, it is added to the set of support points. 
One limitation of ARS is that it can be applied only  with (univariate) log-concave target densities.\footnote{The possibility of applying ARS for drawing for multivariate densities depends on the ability of constructing a sequence of non-parametric proposal pdfs in higher dimensions. See, for instance, the piecewise constant construction in \citep{MartinoIA2RMS15} as a simpler alternative procedure.} For this reason, several extensions have been proposed \citep{Hoermann95,Hirose05,Evans98,Gorur08rev,MartinoStatCo10}, even mixing with MCMC techniques \citep{Gilks95,Sticky13,MartinoIA2RMS15}. A related RS-type method, automatic but non-adaptive, that employs a piecewise constant construction of the proposal density obtained with a pruning  of the initial nodes, has been suggested in \citep{FUSS15}. Another variant has been provided in \citep{PARS}.

In this work, we focus on the computational cost required by ARS. The ARS algorithm obtains high acceptance rates improving the proposal function, which becomes closer and closer to target function. Hence, this enhancement of the acceptance rate is obtained building more complex proposals, which become more computational demanding. The overall time of ARS depends on both the acceptance rate and the time required for sampling from the proposal pdf. The computational cost of ARS remains bounded since the probability of updating the proposal pdf, $P_t$, vanishes to zero as the number of iterations $t$ grows. 
 However, for a finite $t$, there is always a positive probability $P_t>0$ of improving the proposal function, producing an increase of the acceptance rate. This enhancement of the acceptance rate could not balance out the increase of the time required for drawing from the new updated proposal function. Namely, if the acceptance rate is enough close to 1, a further improvement of the proposal function could become prejudicial. 

Thus, we propose a novel ARS scheme, called Cheap Adaptive Rejection Sampling (CARS), employing a fixed number of nodes, i.e., the computational effort required for sampling from the proposal remains constant, selected in advance by the user. The new technique is able to increase the acceptance rate on-line in the same fashion of the standard ARS method, improving adaptively the location of the support points. The configuration of the nodes converges to the best possible distribution which maximizes the acceptance rate achievable with a fixed number of support points. Clearly, the maximum obtainable acceptance rate with CARS is always smaller than 1, in general. However, for large value of required samples, the CARS algorithm is faster than ARS for generating independent samples from the target, as shown the numerical simulations.     

\section{Adaptive Rejection Sampling}
\label{sec:ars}

\begin{figure}[htb]
\centerline{
\includegraphics[width=6cm]{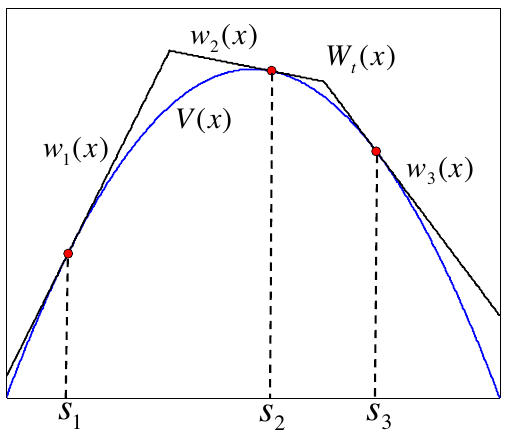}
 		}
\caption{Example of construction of the piecewise linear function $W_t(x)$ (black line) with $m_t=3$ support points, such that $W_t(x)\geq V(x)$ (where $V(x)$ is shown in blue line). The support points, $s_1$, $s_2$ and $s_3$ are depicted with circles.}
\label{FigARS}
\end{figure}

We denote the target density as 
\begin{equation}
\bar{\pi}(x)=\frac{1}{c_\pi} \pi(x)=\frac{1}{c_\pi}\exp\big(V(x)\big), \quad x\in \mathcal{X}\subseteq \mathbb{R},
\end{equation}
with $V(x)=\log[\pi(x)]$ and $c_\pi=\int_{\mathcal{X}} \pi(x)dx$. The adaptive proposal pdf is denoted as 
\begin{equation}
\bar{q}_t(x)= \frac{1}{c_t} q_t(x),
\end{equation}
where $c_t=\int_{\mathcal{X}} q_t(x)dx$, and $t\in \mathbb{N}$. In order to apply rejection sampling (RS), it is necessary to build $q_t(x)$ as an envelope function of $\pi(x)$, i.e.,
\begin{equation}
q_t(x)\geq \pi(x), \quad \mbox{ or } \quad  W_t(x)\geq V(x),  
\end{equation}
where $W_t(x)=\log [q_t(x)]$, for all $x\in \mathcal{X}$ and $t\in \mathbb{N}$. As a consequence, it is important to observe that 
\begin{equation}
c_t\geq c_{\pi}, \qquad \forall t\in \mathbb{N}.
\end{equation}
Let us assume that $V(x) = \log \pi(x)$ is concave, and we are able to evaluate the function $V(x)$ and its first derivative $V'(x)$.\footnote{The evaluation of $V'(x)$ is not strictly necessary, since the function $q_t(x)$ can also construct using a derivative-free procedure (e.g., see \citep{Gilks92derfree} or the piecewise constant construction in \citep{MartinoIA2RMS15}). For the sake of simplicity, we consider the construction involving tangent lines.}
The adaptive rejection sampling (ARS) technique \citep{Gilks92derfree,Gilks92} considers a set of {\it support points} at the $t$-th iteration,
\begin{equation}
	\mathcal{S}_t = \{s_1, s_2,\ldots, s_{m_{t}}\} \subset \mathcal{X},
\end{equation}
such that $s_1 < \ldots < s_{m_{t}}$ and $m_t=|\mathcal{S}_t|$, for constructing the envelope function $q_t(x)$ in a non-parametric way.   We denote as $w_i(x)$ as the straight line tangent to $V(x)$ at $s_i$ for $i=1,\ldots,m_t$. Thus, we can build a piecewise linear function,
\begin{equation}
	W_t(x) = \min[w_1(x), \ldots, w_{m_t}(x)], \quad x\in \mathcal{X}.
\label{EqWtARS}
\end{equation}
Hence, the proposal pdf defined as $\bar{q}_t(x)\propto q_t(x)=\exp(W_t(x))$, is formed by exponential pieces in such a way that $W_t(x) \geq V(x)$, so that $q_t(x)\geq \pi(x)$, when $V(x)$ is concave (i.e., $\pi(x)$ is log-concave). Figure \ref{FigARS} depicts an example of 
piecewise linear function $W_t(x)$ built with $m_t=3$ support points.

\begin{table}[!t]
	\centering
	\caption{Adaptive Rejection Sampling (ARS) algorithm.}
    \begin{tabular}{|p{0.95\columnwidth}|}
		\hline

\textbf{Initialization:}

\begin{itemize}

\item[1.] Set $t=0$ and $n=0$. Choose an initial set $\mathcal{S}_0=\{s_1,\ldots,s_{m_0}\}$.

\end{itemize}

\textbf{Iterations (while $\boldsymbol{n < N}$):}

\begin{itemize}

\item[2.]  Build the proposal $q_t(x)$, given the set of support points $\mathcal{S}_t=\{s_1,\ldots,s_{m_t}\}$, according to Eq. \eqref{EqWtARS}.

\item[3.]  Draw $x' \sim \bar{q}_t(x) \propto q_t(x)$ and $u' \sim \mathcal{U}([0,1])$.

\item[4.] If $u' > \frac{\pi(x')}{q_t(x')}$, then reject $x'$, update 
$$\mathcal{S}_{t+1}=\mathcal{S}_{t}\cup\{x'\},$$ 
and set  $t=t+1$. Go back to step 2. 

\item[5.] If $u' \leq\frac{p(x')}{\pi_t(x')}$, then accept $x'$, setting $x_n=x'$.

\item[6.] Set $\mathcal{S}_{t+1}=\mathcal{S}_{t}$, $t=t+1$, $n=n+1$ and return to step 2.
\end{itemize}
\textbf{Outputs:} The $N$ accepted samples $x_1,\ldots,x_N$.
\\
	\hline
	\end{tabular}
	\label{tab:ars}
\end{table}

Table \ref{tab:ars} summarizes the ARS algorithm for drawing $N$ independent samples from $\bar{\pi}(x)$. At each iteration $t$, a sample $x'$ is drawn from $\bar{q}_t(x)$ and accepted with probability $\frac{\pi(x')}{q_t(x')}$, otherwise is rejected.  Note that a new point is added to the support set $\mathcal{S}_t$ whenever it is rejected in the RS test improving the construction of $q_t(x)$. Clearly, denoting as $T$ the total number of iterations of the algorithm, we have always $T\geq N$ since several samples are discarded.
%
\section{Computational cost of ARS}
The computational cost of an ARS-type method, in a specific iteration $t$, depends on two elements:
\begin{enumerate}
\item The number of samples accepted in RS test (averaged over different runs), i.e., the acceptance rate. 
\item The computational effort required for sampling from $q_t(x)$.
\end{enumerate}
We desire that the acceptance rate is close to 1 and, simultaneously, that the spent time required for drawing from $q_t(x)$ is small. In general, there exists a trade-off since an increase of the acceptance rate requires the use of a more complicated proposal density $q_t(x)$. ARS is an automatic procedure which provides a possible compromise. Below, we analyze some important features of a standard ARS scheme.

\subsection{Acceptance rate}
The averaged number of accepted samples, i.e., the acceptance rate, is 
\begin{equation}
 \eta_t = \int  \frac{\pi(x)}{q_t(x)}  \bar{q}_t(x)dx= \frac{c_\pi}{c_t},
	\end{equation}
that is $0\leq\eta_t\leq 1$ since $c_t\geq c_\pi$, $\forall t\in \mathbb{N}$, by construction. Note that in an ARS scheme, $\eta_t$ varies from a realization to other since $c_t$ is different due to the set $\mathcal{S}_t$ and, as a consequence, $q_t$ are randomly constructed at each run.\footnote{In the following, we denote as $E[\eta_t]$ the acceptance rate, at the $t$-th iteration, averaged over several (theoretically infinite) runs.} Defining the $L_1$ distance between $\pi_t(x)$ and $p(x)$ as
\begin{equation}
	D(q_t,\pi) = \|q_t(x)-\pi(x)\|_1 = \int_{\mathcal{X}}{|q_t(x) - \pi(x)| dx},
\label{es:distanceL1}
\end{equation}
ARS ensures that $D(q_t,\pi) \to 0$ when $t \to \infty$, and as a consequence $c_t\rightarrow c_\pi$. Thus, $ \eta_t$ tends to one as $t \rightarrow \infty$. Indeed, as $\eta_t \rightarrow 1$, ARS becomes virtually an exact sampler after a some iterations.

\subsection{Drawing from the proposal pdf}
Let us denote the exponential pieces as
\begin{equation}
h_i(x)=e^{w_i(x)},  \quad i=1,\ldots,N,
\end{equation}
so that 
$$
q_t(x)=h_i(x), \quad \mbox{ for } \quad x\in \mathcal{I}_i=(e_{i-1}, e_{i}], \quad  i=1,\ldots,N,
$$
where $e_{i}$ is the intersection point between the straight lines $w_{i}(x)$ and $w_{i+1}(x)$, for $i=2,\ldots,N-1$, and $e_{0}=-\infty$ and $e_{N}=+\infty$ (if $\mathcal{X}=\mathbb{R}$).
Thus, for drawing a sample $x'$ from $\bar{q}_t(x)=\frac{1}{c_t} q_t(x)$, we need to:
\begin{enumerate}
\item Compute analytically the area $A_i$ below each exponential piece, i.e., $A_i=\int_{ \mathcal{I}_i} h_i(x)dx$ and obtain the normalized weights
\begin{equation}
\rho_i=\frac{A_i}{\sum_{n=1}^N A_n}= \frac{A_i}{c_t},
\end{equation}
where we have observed that $c_t=\sum_{n=1}^N A_n=\int_{\mathcal{X}} q_t(x)dx$.
\item Select an index $j^*$ (namely, one piece) according to the probability mass $\rho_i$, $i=1,\ldots,N$.
\item Draw  $x'$ from $h_{j^*}(x)$ restricted within the domain $\mathcal{I}_{j^*}=(e_{j^*-1}, e_{j^*}]$, and zero outside (i.e., from a truncated exponential pdf).
\end{enumerate}
Observe that, at step 2, a multinomial sampling is required. It is clear that the computational cost for drawing  one sample from $q_t(x)$ increases as the number of pieces grows or, equivalently, the number of support points grows. Fortunately, the computational cost in ARS is automatically controlled by the algorithm, since the probability of adding a new support point
\begin{equation}
P_t = 1-\eta_t = \frac{1}{c_t}D(q_t,\pi),
\end{equation}
 tends to zero as $t \rightarrow \infty$, since the distance in Eq. \eqref{es:distanceL1} vanishes to zero, i.e., $D(q_t,\pi) \to 0$.

\section{ARS with fixed number of support points}
 We have seen that the probability of adding a new support point $P_t$ vanishes to zero as $t \rightarrow \infty$. However, for a finite $t$, we have always a positive probability $P_t>0$ of adding a new point (although small), so that a new support point could be incorporated producing an increase of the acceptance rate. After a certain iteration $\tau$, i.e., $t>\tau$, this improvement of the acceptance rate could not balance out the increase of the time required for drawing from the proposal, due to the addition of the new point. Namely, if the acceptance rate is enough close to 1, a further addition of a support point could slow down the algorithm, becoming prejudicial. 

In this work, we provide an alternative adaptive procedure for ARS, called {\it Cheap Adaptive Rejection Sampling} (CARS), which uses a fixed number of support points. When a sample is rejected, a test for swapping the rejected sample with the closest support point within $\mathcal{S}_t$ is performed, so that the total number of points remains constant. Unlike in the standard ARS method, in the new adaptive scheme the test is deterministic. The underlying idea is based on the following observation. The standard ARS algorithm yields a decreasing sequence of normalizing constants $\{c_t\}_{t\in\mathbb{N}}$ of the proposal pdf converging to $c_\pi=\int_{\mathcal{X}} \pi(x)dx$, i.e.,
\begin{equation}
c_0\geq c_1 \ldots \geq c_t \ldots \geq c_{\infty}=c_\pi.
\end{equation}
Clearly, since the acceptance rate is $\eta_t=\frac{c_\pi}{c_t}$ this means that $\eta_t\rightarrow 1$.
In CARS, we provide an alternative way for producing this  decreasing sequence of normalizing constants $\{c_t\}$. Indeed, an exchange between two points is accepted if it produces a reduction in the normalizing constant of the corresponding proposal pdf. More specifically, consider the set  
$$
\mathcal{S}_t=\{s_1,s_2,\ldots,s_M\},
$$
contained $M$ support points. When a sample $x'$ is rejected in the RS test, the closest support point $s^*$ in $\mathcal{S}_t$ is obtained, i.e.,
$$
s^*=\arg\min_{s_i\in\mathcal{S}_t} |s_i -x'|.
$$
We recall that we denote with $q_t(x)$ the proposal pdf built using $\mathcal{S}_t$ and with $c_t$ its normalizing constant. Then, we consider a new set 
\begin{equation}
\mathcal{G}=\mathcal{S}_t\cup \{x'\} \backslash \{s^*\},
\end{equation}
namely, including $x'$ and removing $s^*$. We denote with $g(x)$ the proposal built using the alternative set of support points $\mathcal{G}$, and $c_g=\int_{\mathcal{X}} g(x) dx$. If 
$$
c_g<c_t,
$$
then the swap is accepted, i.e., we set $\mathcal{S}_{t+1}=\mathcal{G}$ for the next iteration, otherwise the set remains unchanged, $\mathcal{S}_{t+1}=\mathcal{S}_t$. The complete algorithm is outlined in Table \ref{tab:cars}. Note that $c_t$ is always computed (in any case, for both ARS and CARS) at the step 3, for sampling from $q_t(x)$. Furthermore observe that, after the first iteration, step 2 can be skipped since the new proposal pdf $q_{t+1}(x)$ has been already constructed in the previous iteration, i.e., $q_{t+1}(x)=q_{t}(x)$, or at step 4.3, i.e., $q_{t+1}(x)=g(x)$. 

Therefore, with the CARS algorithm, we obtain again a decreasing sequence of $\{c_t\}_{t\in\mathbb{N}}$
$$
c_0\geq c_1 \ldots \geq c_t \ldots \geq c_{\infty},
$$
but $c_{\infty}\neq c_\pi$ so that $\eta_t \rightarrow \eta_{\infty} <1$, in general. The value $\eta_{\infty}$ is the highest acceptance rate that can be obtained with $M$ support points, given the target function $\pi(x)$. Therefore, CARS yields a sequence of sets $\mathcal{S}_1,\ldots,\mathcal{S}_t,\ldots$ that converges to the stationary set $\mathcal{S}_\infty$ containing the best configuration of $M$ support points for maximizing the acceptance rate, when the target function is $\pi(x)$ and given a specific construction procedure for the proposal $q_t(x)$.\footnote{The best configuration $\mathcal{S}_\infty$  depends on the specific construction procedure employed for building the sequence of proposal functions $q_1,q_2\ldots,q_t,\ldots$} 

\begin{table}[!t]
	\centering
	\caption{Cheap Adaptive Rejection Sampling (CARS) algorithm.}
    \begin{tabular}{|p{0.95\columnwidth}|}
		\hline

\textbf{Initialization:}

\begin{itemize}

\item[1.] Set $t=0$ and $n=0$. Choose a value $M$ and an initial set $\mathcal{S}_0=\{s_1,\ldots,s_{M}\}$.

\end{itemize}

\textbf{Iterations (while $\boldsymbol{n < N}$):}

\begin{itemize}

\item[2.]  Build the proposal $q_t(x)$, given the current set $\mathcal{S}_t$, according to Eq. \eqref{EqWtARS} or other suitable procedures.

\item[3.]  Draw $x' \sim \bar{q}_t(x) \propto q_t(x)$ and $u' \sim \mathcal{U}([0,1])$.

\item[4.] If $u' > \frac{\pi(x')}{q_t(x')}$, then reject $x'$ and:
\begin{itemize}
\item[4.1] Find the closest point $s^*$  in $\mathcal{S}_t$,
$$
s^*=\arg\min_{s_i\in\mathcal{S}_t} |s_i -x'|.
$$
\item[4.2] Build the alternative proposal $g(x)$ based on the set of points
$$\mathcal{G}=\mathcal{S}_t\cup \{x'\} \backslash \{s^*\}$$
 and compute $c_g=\int_{\mathcal{X}} g(x) dx$.
\item[4.3] If $c_g<c_t$, set $\mathcal{S}_{t+1}=\mathcal{G},$ otherwise, if $c_g\geq c_t$, set $\mathcal{S}_{t+1}=\mathcal{S}_{t}$. Set  $t=t+1$ and go back to step 2. 
\end{itemize}
\item[5.] If $u' \leq\frac{p(x')}{\pi_t(x')}$, then accept $x'$, setting $x_n=x'$.

\item[6.] Set $\mathcal{S}_{t+1}=\mathcal{S}_{t}$, $t=t+1$, $n=n+1$ and return to step 2.
\end{itemize}
\textbf{Outputs:} The $N$ accepted samples $x_1,\ldots,x_N$.
\\
	\hline
	\end{tabular}
	\label{tab:cars}
\end{table}

In Table \ref{tab:cars}, the possibility of changing the current set $\mathcal{S}_t$ is given only if $x'$ is rejected in the RS test. Namely, only a subset of all the generated samples $x'$ from the proposal $\bar{q}_t$ are considered as a possible new support point. However, sampling and adaptation could be completely divided. For instance, the alternative proposal pdf $g(x)$ could be constructed (and then $c_g$ could be computed) considering any sample $x'$ generated by $\bar{q}_t(x)$ at Step 3 of the algorithm (not only the rejected ones). In this case, Steps 4.1, 4.2, 4.3 of Table \ref{tab:cars} would be performed at each iteration, so that the corresponding algorithm would be probably slowed down with respect to version of CARS described in Table \ref{tab:cars}.
\newline
{\bf About the choice of $M$.} The user can choose the number of nodes $M$ according to the available computational resources. Note that, when $M$ grows, the computational effort for sampling from ${\bar q}_t(x)$ increases but, at the same time, a greater acceptance rate can be obtained. This trade-off explains the possible existence of an optimal value $M^*$, as shown in Figure \ref{Fig1_Ex2}(b). The optimal value $M^*$(when exists) depends on the target pdf and the capability of the employed processor/machine.

\section{Numerical simulations}

In order to show the capability of the novel technique, we compare the performance the standard ARS and CARS methods consider two well-known log-concave target densities, Gaussian and Gamma pdfs, as typical examples of log-concave, symmetric and skewed distributions, respectively. 

\subsection{Gaussian distribution}
We consider a Gaussian density as (typical) log-concave target pdf and test both ARS and CARS. Namely, we consider
$$
\bar{\pi}(x)\propto \pi(x)=\exp\left(-\frac{x^2}{2\sigma^2}\right), \quad x\in \mathbb{R},
$$
with $\sigma^2=\frac{1}{2}$. We compare ARS and CARS in terms of the time required for generating $N\in\{5000,10000,50000\}$ samples. In all cases and both techniques, we consider a initial set of support points $\mathcal{S}_0=\{s_1,\ldots,s_{m_0}\}$ with cardinality $m_0=|\mathcal{S}_0|\in\{3,5,10\}$ (clearly, $M=m_0$ in CARS) where the initial points are chosen uniformly in $[-2,2]$ at each simulation, i.e., $s_i\sim\mathcal{U}([-2,2])$.\footnote{Clearly, the configurations of either all negative or all positive are discarded since they yield improper proposal pdf by construction. } 

\begin{figure}[htb]
\centerline{
\subfigure[]{\includegraphics[width=7cm]{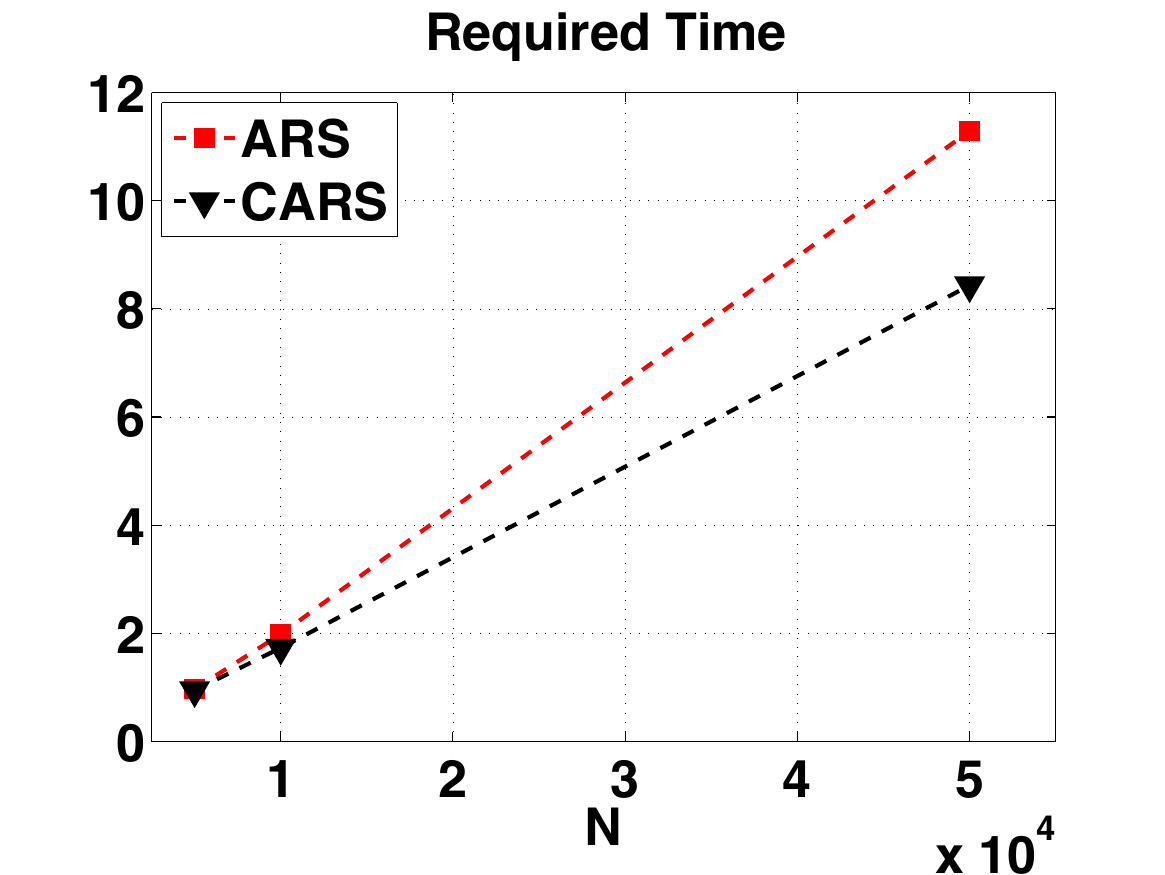}}
\subfigure[]{\includegraphics[width=7cm]{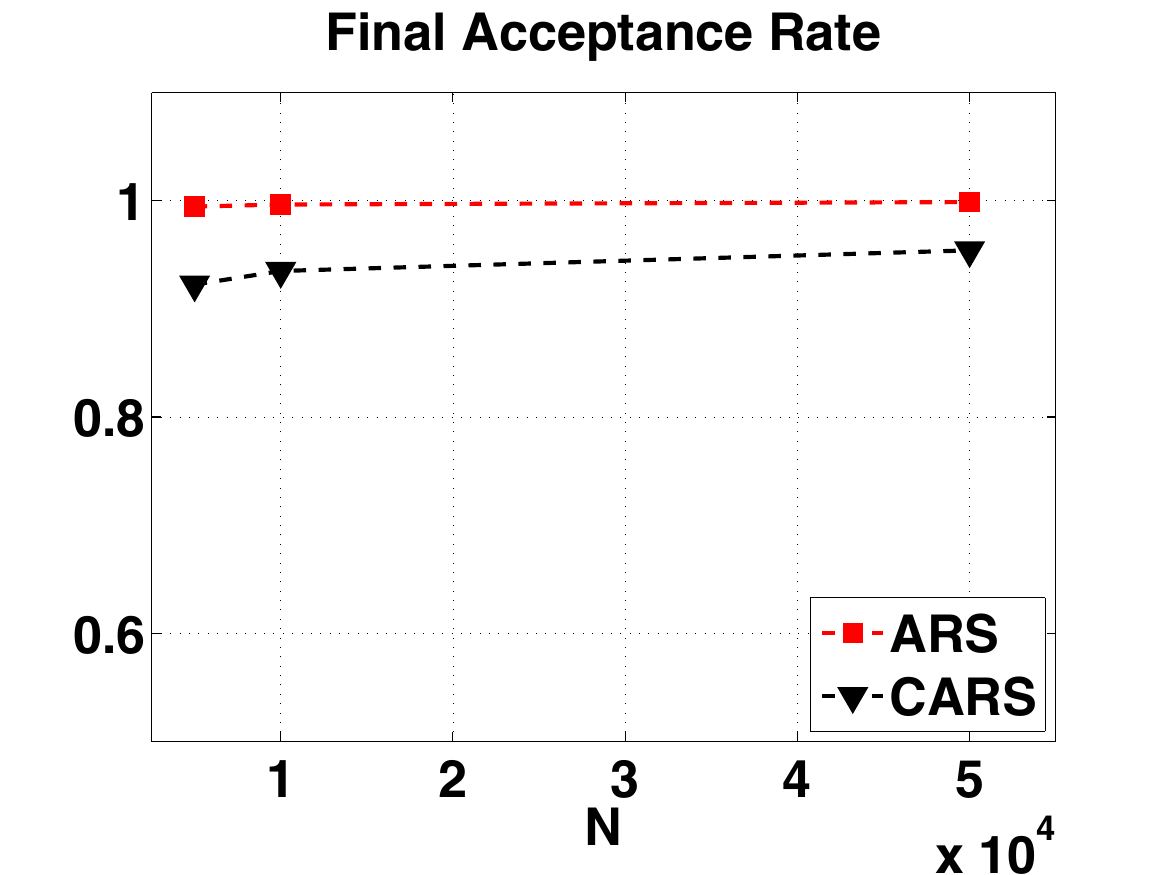}}
}
\centerline{
\subfigure[]{\includegraphics[width=7cm]{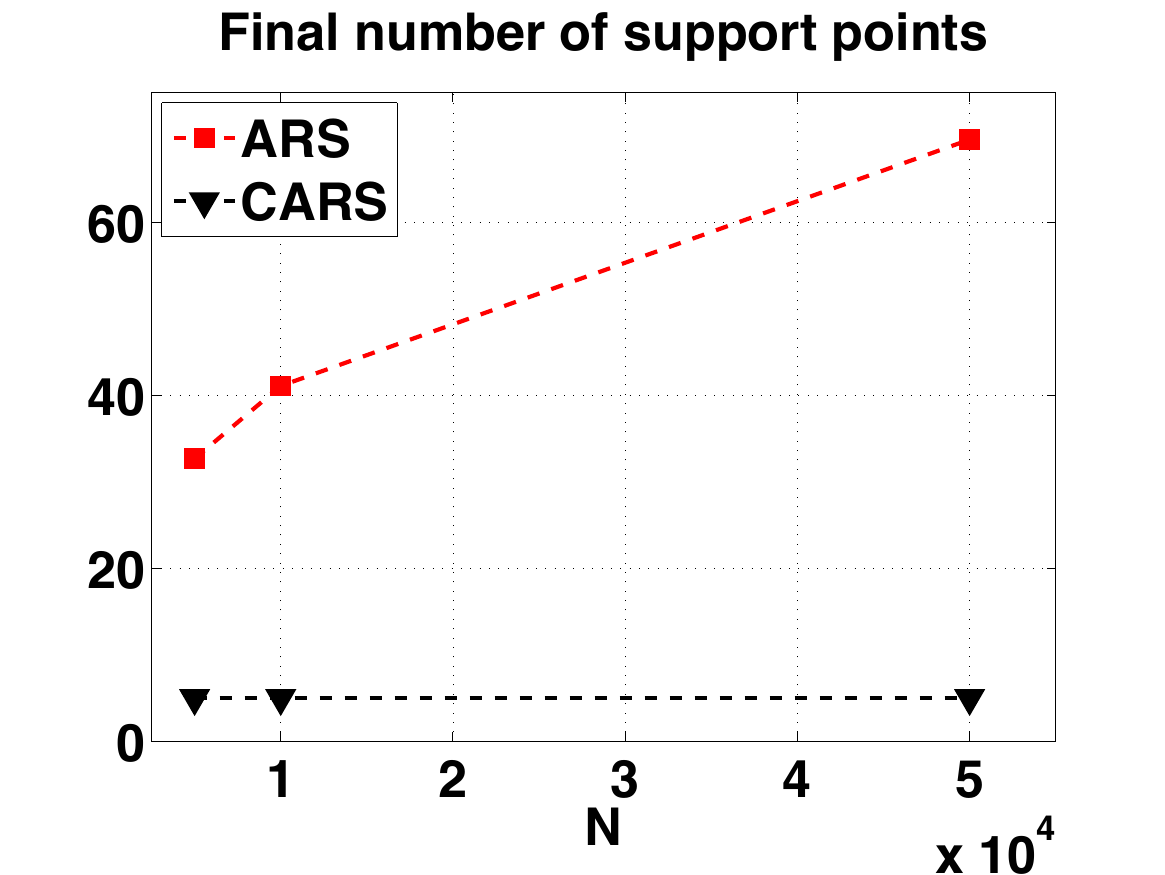}}
 		}
\caption{{\bf (a)} Spent time (normalized w.r.t. the time required by ARS with $N=5000$, $m_0=3$), {\bf (b)} final acceptance rate, and {\bf (c)} final number of support points,
as function of the number $N$ of drawn samples, for ARS (squares) and CARS (triangles). 
}
\label{FigSimu1}
\end{figure}

\begin{figure}[htb]
\centerline{
\subfigure[]{\includegraphics[width=7cm]{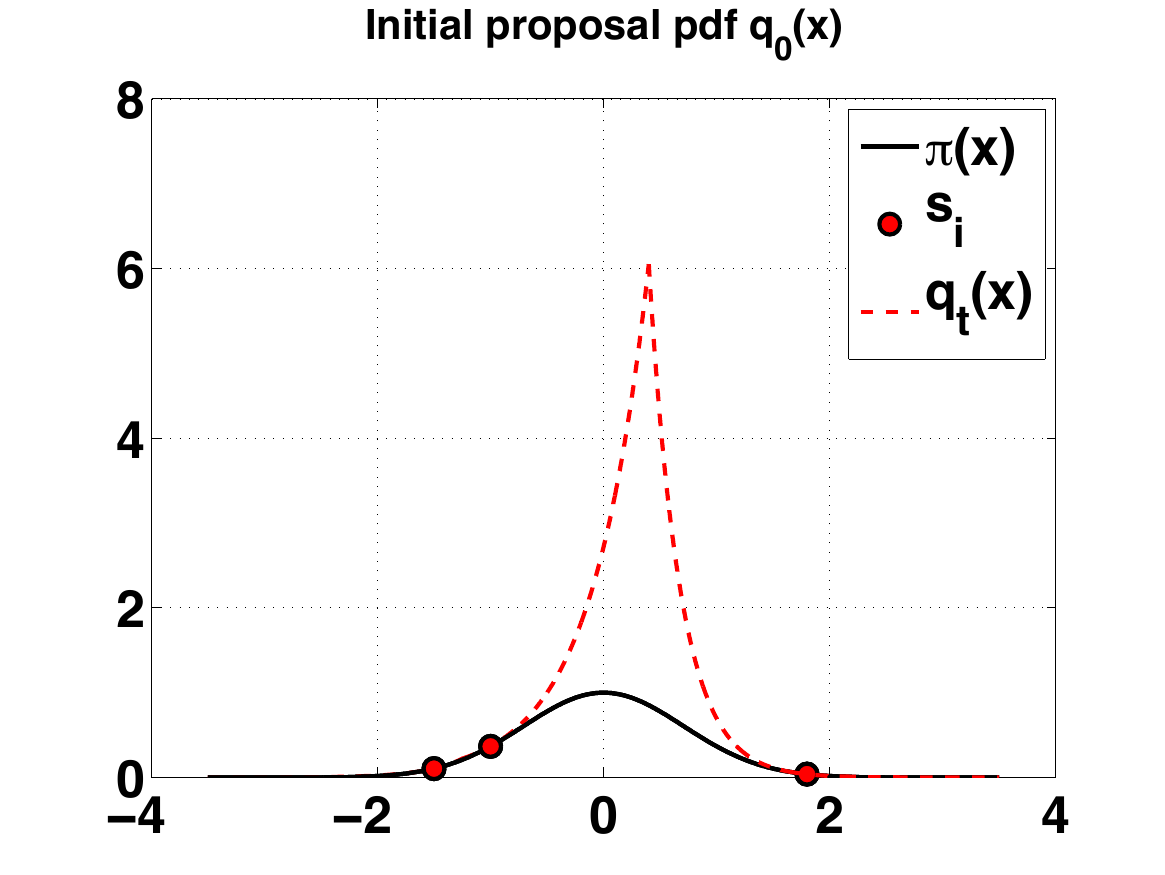}}
\subfigure[]{\includegraphics[width=7cm]{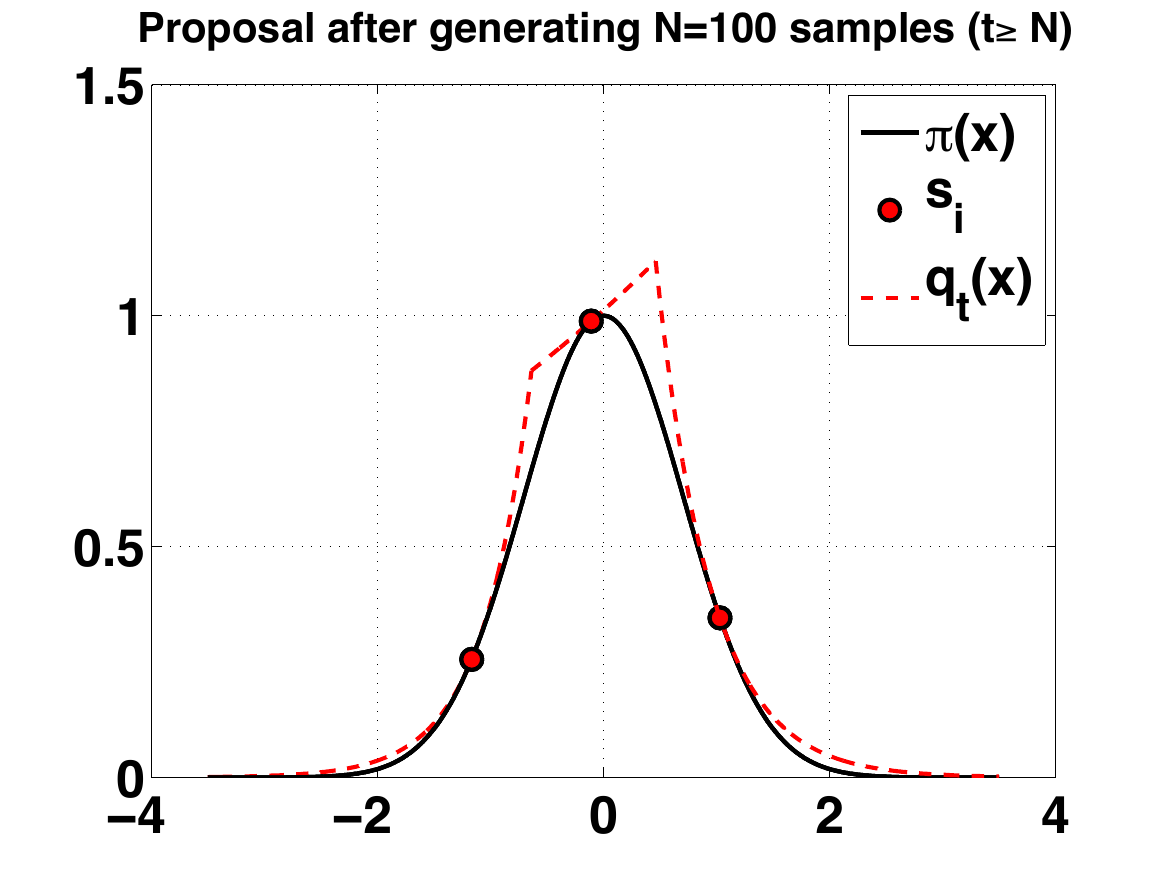}}
}
\centerline{
\subfigure[]{\includegraphics[width=7cm]{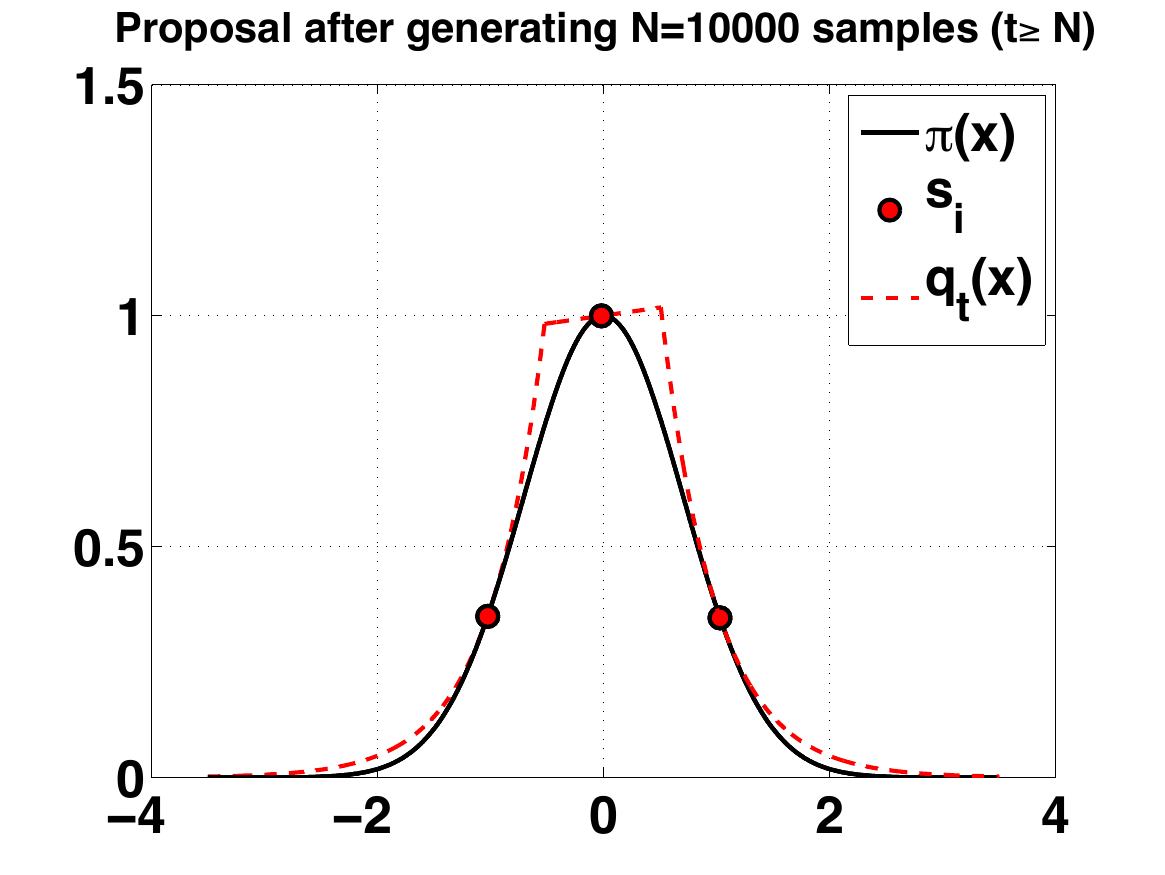}}
 		}
\caption{Example of sequence of proposal pdfs obtained by CARS, starting with $S_0=\{-1.5,-1,1.8\}$. We can observe that the $L_1$ distance $D(q_t,\pi)$ is reduced progressively. The proposal function $q_t(x)$ is depicted with dashed line,  the target function $\pi(x)$ with solid line and the support points with circles. The configuration of the nodes in figure {\bf (c)} is $S_t=\{ -1.0261, -0.0173, 1.0305\}$ with $t\geq N=10^4$. The optimal configuration with $3$ nodes and $ \pi(x)=\exp\left(-x^2\right)$ is $S_\infty=\{ -1, 0, 1\}$.
}
\label{FigSimu2}
\end{figure}

We run $500$ independent simulations for each case and compute the required time for generating $N$ samples (using a Matlab code), the averaged number of final support points (denote as $E[m_T]$) and the acceptance rate reached in the final iteration (denoted as $E[\eta_T]$; averaged  over the 500 runs), for both techniques. Table \ref{table1res} shows the results. The time is normalized with respect to (w.r.t.) the time spent by ARS with $N=5000$, $m_0=|\mathcal{S}_0|=3$. 
The results show that CARS is always faster than ARS. We can observe that both methods obtain acceptance rates close to 1. CARS reaches acceptance rates always greater of $0.87$ using only $3$ nodes. CARS obtains an acceptance rate $E[\eta_T]$ more than $0.98$ employing only $10$ nodes and after generating $N=5000$ independent samples. Fig. \ref{FigSimu1} depicts the spent time, the final acceptance rate and the final number of nodes, as function of number $N$ of generated samples. We can observe that CARS is significantly faster than ARS when $N$ grows, owing to ARS yields a sensible increase of the number of support points that corresponds to an infinitesimal increase of the acceptance rate, whereas in CARS the number of nodes remains constant.  Figure \ref{FigSimu2} shows a sequence of proposal pdfs constructed by CARS, using $3$ nodes and starting with $S_0=\{-1.5,-1,1.8\}$. The  $L_1$ distance $D(q_t,\pi)$ is reduced progressively and the acceptance rate improved. The final set of support point is $S_t=\{ -1.0261, -0.0173, 1.0305\}$, close to the optimal one $S_\infty=\{ -1, 0, 1\}$.

\begin{table}[!htb]
\caption{Results as function of the desired number of samples $N$ and the cardinality $|\mathcal{S}_0|$ of the initial set of support points $\mathcal{S}_0$.  We show the normalized spent time, the averaged final number of support points, $E[m_T]$, and the averaged final acceptance rate, $E[\eta_T]$.  
 }
\label{table1res}
\footnotesize
\begin{center}
\begin{tabular}{|c|c|c|c|c|}
\hline
{\bf Scheme} & $N$ & $|\mathcal{S}_0|=3$ & $|\mathcal{S}_0|=5$  & $|\mathcal{S}_0|=10$    \\
\hline
\hline
  \multirow{ 3}{*}{ARS}   & \multirow{ 3}{*}{$5000$} & Time=$1$ &  Time=$0.9709$  & Time=$ 0.9801$  \\
  & & $E[\eta_T]=0.9942$ & $E[\eta_T]=0.9945$ & $E[\eta_T]=0.9952$    \\
  &  & $E[m_T]=32.36$    & $E[m_T]=32.69$  &  $E[m_T]=34.17$ \\
\hline
\hline
   \multirow{3}{*}{CARS}   & \multirow{ 3}{*}{$5000$} & Time=$0.9599$ &  Time=$0.9477$  & Time=$0.9694$  \\
  & & $E[\eta_T]=0.8721$ & $E[\eta_T]= 0.9224$ & $E[\eta_T]=0.9556$    \\
  &  & $E[m_T]=M=3$    & $E[m_T]=M=5$  &  $E[m_T]=M=10$ \\
 \hline
  \hline
   \multirow{ 3}{*}{ARS}   & \multirow{ 3}{*}{$10000$} &  Time=$2.2843$ &  Time=$1.9862$  & Time=$1.9983$  \\
  & & $E[\eta_T]=0.9963$ & $E[\eta_T]=0.9964$ & $E[\eta_T]=0.9968$    \\
  &  & $E[m_T]=40.60$    & $E[m_T]=41.09$  &  $E[m_T]=42.16$ \\
\hline
\hline
   \multirow{3}{*}{CARS}   & \multirow{ 3}{*}{$10000$} & Time=$1.9716$ &  Time=$ 1.7311 $  & Time=$1.8969$  \\
  & & $E[\eta_T]=0.8784$ & $E[\eta_T]=0.9350$ & $E[\eta_T]=0.9631$    \\
  &  & $E[m_T]=M=3$    & $E[m_T]=M=5$  &  $E[m_T]=M=10$ \\
 \hline
 \hline
  \multirow{ 3}{*}{ARS}   & \multirow{ 3}{*}{$50000$} & Time=$11.2196$  &  Time=$11.2887$  & Time=$11.7599$  \\
  & & $E[\eta_T]=0.9987$ & $E[\eta_T]=0.9987$ & $E[\eta_T]= 0.9988$    \\
  &  & $E[m_T]=68.63$    & $E[m_T]=69.56$  &  $E[m_T]=70.09$ \\
\hline
\hline
   \multirow{3}{*}{CARS}   & \multirow{ 3}{*}{$50000$} & Time=$8.7756 $ &  Time=$8.4322$  & Time=$9.0704$  \\
  & & $E[\eta_T]=0.8855$ & $E[\eta_T]=0.9540$ & $E[\eta_T]=0.9861$    \\
  &  & $E[m_T]=M=3$    & $E[m_T]=M=5$  &  $E[m_T]=M=10$ \\
 \hline

\end{tabular}
\end{center}
\end{table}%

\subsection{Gamma distribution}
In this section, we consider a Gamma density
$$
\bar{\pi}(x)\propto \pi(x)=x^{r-1}\exp\left(-\frac{x}{a}\right), \quad x\in \mathbb{R},
$$
with $r=2$ and $a=2$.  In all the experiments, we consider an initial set of support points $\mathcal{S}_0=\{s_1=0.01,\ldots,s_i,\ldots,s_{m_0}=4\}$ with cardinality $m_0=|\mathcal{S}_0|$, where $s_i\sim\mathcal{U}([0,4])$, with $i=2,\ldots, m_0-1$. Recall that $M=m_0$ in CARS. We consider different number of desired samples $N\geq 5$, and compute the spent time,  the averaged number of final support points (denote as $E[m_T]$) and the acceptance rate reached in the final iteration (denoted as $E[\eta_T]$) averaged over the 500 independent runs.

\begin{figure}[htb]
\centerline{
\subfigure[]{\includegraphics[width=8cm]{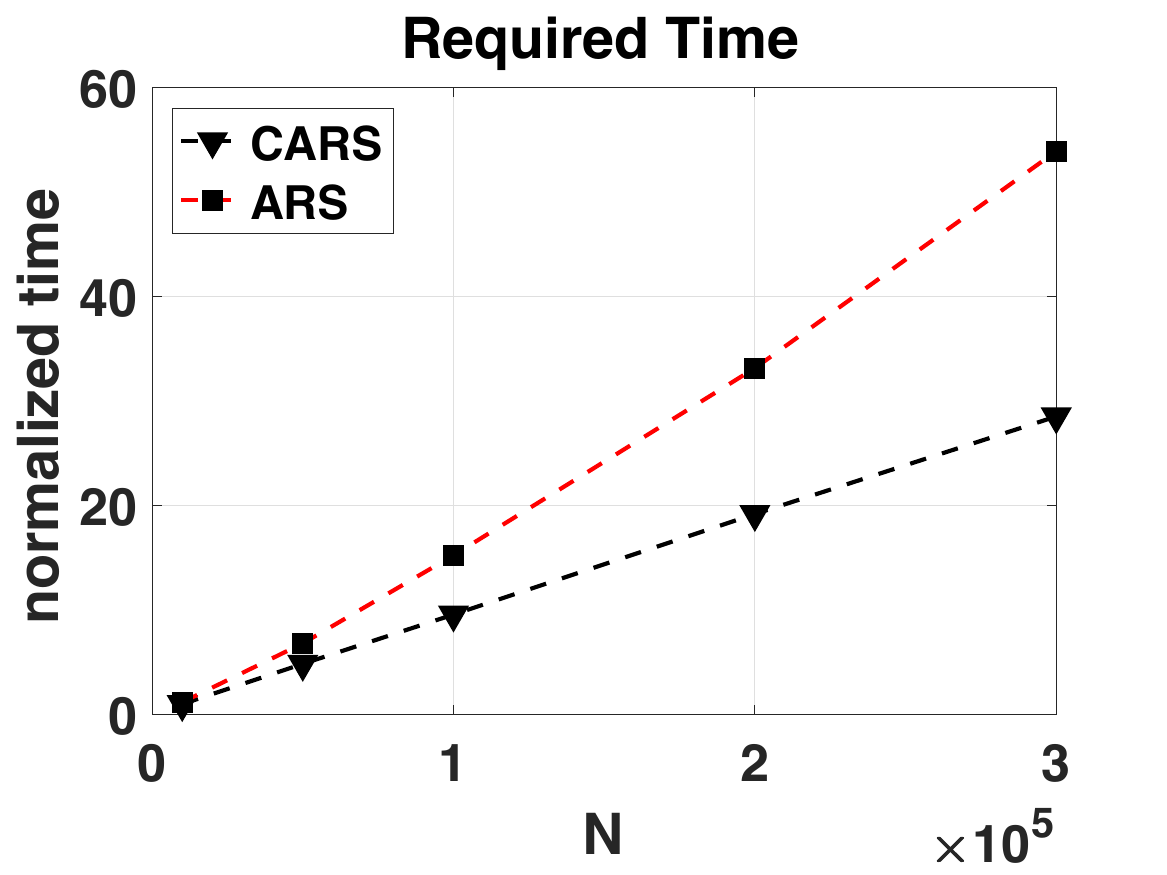}}
\subfigure[]{\includegraphics[width=8cm]{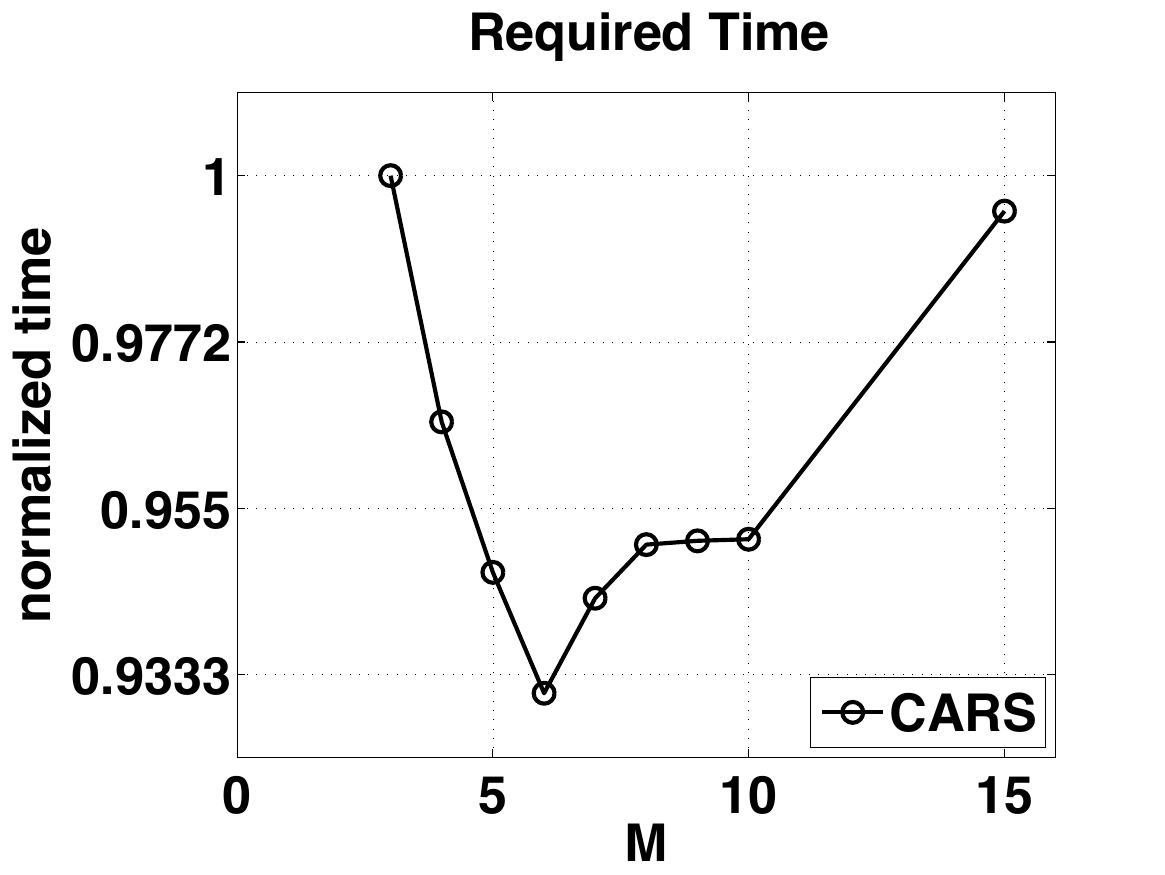}}
 		}
\caption{{\bf (a)} Normalized spent time as function of the number $N$ of drawn samples, for ARS (squares) and CARS (triangles), fixing $m_0=10$ (recall that $M=m_0$ for CARS). The values are normalized w.r.t. the value obtained by CARS with $N=10^4$. {\bf (a)} Spent time as function of the number of nodes $M$ fixing $N=10^5$ (normalized w.r.t. the value obtained by CARS with $M=3$).}
\label{Fig1_Ex2}
\end{figure}

\begin{figure}[htb]
\centerline{
\subfigure[]{\includegraphics[width=8cm]{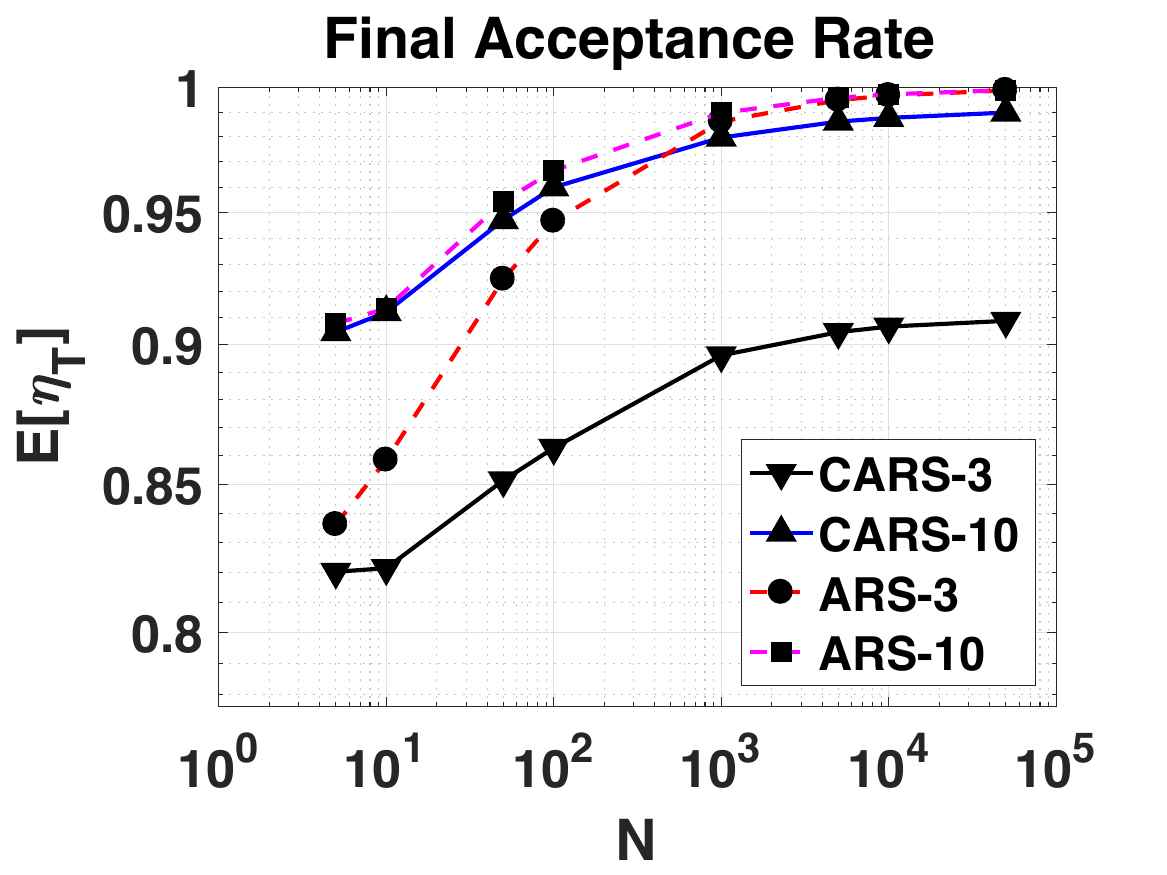}}
\subfigure[]{\includegraphics[width=8cm]{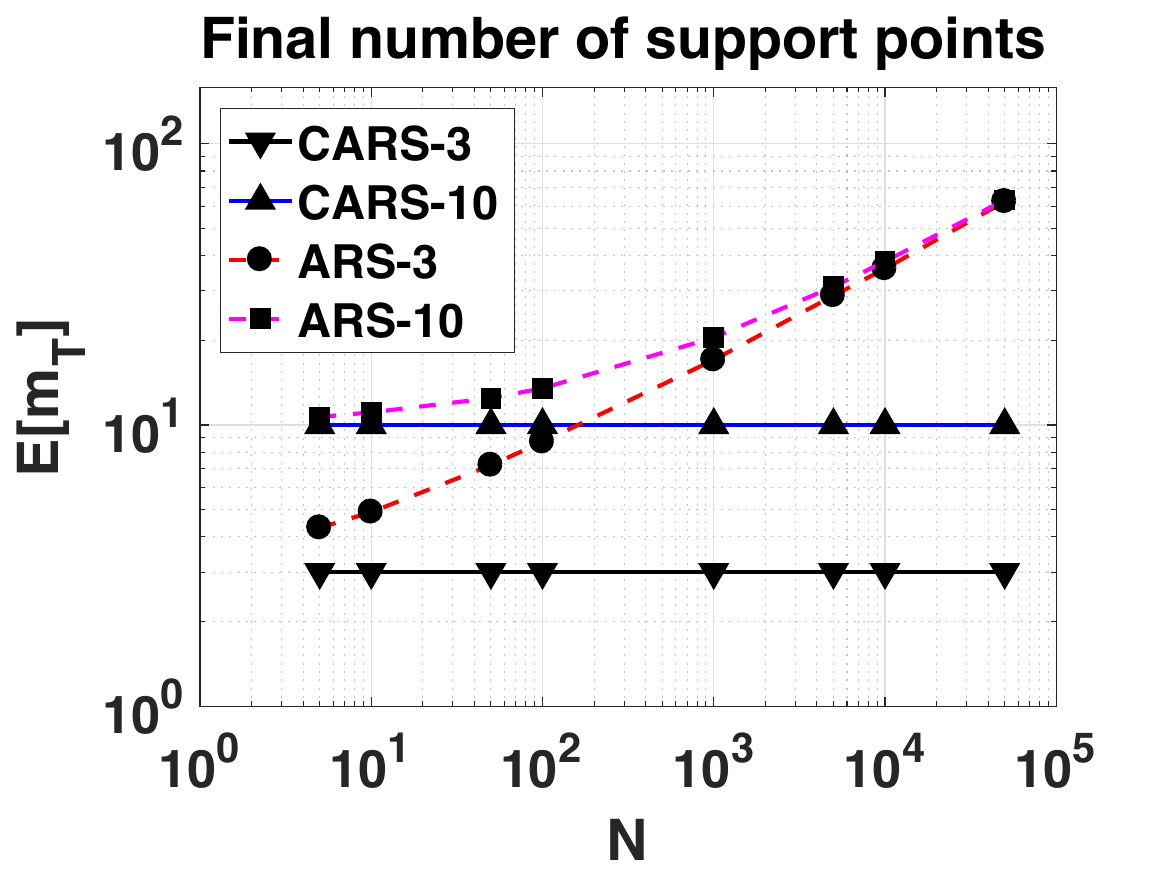}}
 		}
\caption{{\bf (a)} Final averaged acceptance rate $E[\eta_T]$ and {\bf (b)} final number of nodes $E[m_T]$ for ARS (dashed lines) and CARS (solid lines) with $m_0\in\{3,10\}$ (recall $M=m_0$ in CARS), as function of the desired number $N$ of samples.  Both plots are provided in log-log-scale.  }
\label{Fig2_Ex2}
\end{figure}

Figure \ref{Fig1_Ex2}(a) shows the averaged time spent by ARS (with $m_0=10$) and CARS (with $M=10$) as function of the desired number $N$ samples. All the values are normalized w.r.t. the time obtained by CARS with $N =10^4$. Figure \ref{Fig1_Ex2}(b) provides the averaged time values (fixing $N=10^5$) required by CARS as function of $M$ (normalized w.r.t. the value obtained by CARS with $M=3$). We can observe that the time variation is small. However, it seems that an optimal value $M^*$ exists around $M= 6$. Figures \ref{Fig2_Ex2} show the averaged final acceptance rate and final number of nodes in log-log-scale, with $m_0\in\{3,10\}$.



We can observe the number of nodes in the standard ARS increases withe the same speed regardless the initial value $m_0$. Furthermore, CARS with $M=10$ virtually obtains the same curve of acceptance rate than the corresponding standard ARS. We see again that CARS is faster than ARS when $N$ grows, owing to ARS yields a sensible increase of the number of support points that corresponds to an infinitesimal increase of the acceptance rate, whereas in CARS the number of support points remains constant.  

\section{Conclusions}
In this work, we have introduced a novel ARS scheme, the Cheap Adaptive Rejection Sampling (CARS), which employs a fixed number of nodes for the construction of the non-parametric proposal density. As a consequence, the computational effort required for sampling from the proposal remains constant, selected in advance by the user. The new technique is able to increase the acceptance rate on-line in the same fashion of the standard ARS method, improving adaptively the location of the support points. The numerical experiments have shown that,  in order to generate a large number of desired samples, CARS is faster than ARS.

\section{Acknowledgements}
This work has been supported by the Grant 2014/23160-6 of S\~ao Paulo Research Foundation (FAPESP) and by the Grant 305361/2013-3 of National Council for Scientific and Technological Development (CNPq).

\bibliographystyle{unsrtnat}
\bibliography{bibliografia}

\end{document}